\documentclass[aps,prl,preprint,showpacs,preprintnumbers,amsmath,amssymb,superscriptaddress]{revtex4}%

\usepackage{graphicx}%
\usepackage{dcolumn}
\usepackage{amsmath}

\makeatletter
\def\btt#1{\texttt{\@backslashchar#1}}%
\DeclareRobustCommand\bblash{\btt{\@backslashchar}}%
\makeatother

\begin{document}

\preprint{PREPRINT (\today)}

\title{Pressure effects on the transition temperature and the magnetic
field penetration depth in the pyrochlore
superconductor RbOs$_2$O$_6$}

\author{R.~Khasanov}
\affiliation{ Laboratory for Neutron Scattering, ETH Z\"urich and
Paul Scherrer Institut, CH-5232 Villigen PSI, Switzerland}
\affiliation{DPMC, Universit\'e de Gen\`eve, 24 Quai
Ernest-Ansermet, 1211 Gen\`eve 4, Switzerland}
\affiliation{Physik-Institut der Universit\"{a}t Z\"{u}rich,
Winterthurerstrasse 190, CH-8057, Z\"urich, Switzerland}
\author{D.G.~Eshchenko}
\affiliation{Physik-Institut der Universit\"{a}t Z\"{u}rich,
Winterthurerstrasse 190, CH-8057, Z\"urich, Switzerland}
\affiliation{Laboratory for Muon Spin Spectroscopy, PSI, CH-5232
Villigen PSI, Switzerland}
\author{J.~Karpinski}
\affiliation{Solid State Physics Laboratory, ETH 8093 Z\"urich,
Switzerland}
\author{S.M.~Kazakov}
\affiliation{Solid State Physics Laboratory, ETH 8093 Z\"urich,
Switzerland}
\author{N.D.~Zhigadlo}
\affiliation{Solid State Physics Laboratory, ETH 8093 Z\"urich,
Switzerland}
\author{R.~Br\"utsch}
\affiliation{Laboratory for Material Behaviour, Paul Scherrer
Institut, CH-5232 Villigen PSI, Switzerland}
\author{D.~Gavillet}
\affiliation{Laboratory for Material Behaviour, Paul Scherrer
Institut, CH-5232 Villigen PSI, Switzerland}
\author{H.~Keller}
\affiliation{Physik-Institut der Universit\"{a}t Z\"{u}rich,
Winterthurerstrasse 190, CH-8057, Z\"urich, Switzerland}

\begin{abstract}
We report
magnetization measurements under high hydrostatic pressure
in the newly discovered pyrochlore superconductor
RbOs$_2$O$_6$ ($T_c\simeq6.3$~K at $p=0$).
A pronounced and {\it positive} pressure effect (PE) on $T_c$ with
${\rm d} T_c/{\rm d} p =0.090(1)$~K/kbar was observed, whereas no
PE on
the magnetic penetration depth $\lambda$
was detected. The relative pressure shift of $T_c$ [${\rm d} \ln
T_c/{\rm d} p \simeq 1.5$~\%/kbar]
is
comparable with the
highest values obtained for
highly underdoped high-temperature
cuprate superconductors.
Our results
suggest
that RbOs$_2$O$_6$  is
an adiabatic BCS--type superconductor.

\end{abstract}
\pacs{74.70.Dd, 74.62.Fj, 74.25.Ha, 83.80.Fg}

\maketitle

There is an increasing interest to the physics of geometrically
frustrated systems. One of the most remarkable examples is the
observation of bulk superconductivity in pyrochlore oxide
KOs$_2$O$_6$ with the transition temperature $T_c\simeq 9.6$~K
\cite{Yonezawa04}. It was the second compound with pyrochlore
structure, after CdRe$_2$O$_7$ ($T_c\simeq1$~K)
\cite{Hanawa01,Sakai01}, where superconductivity was observed.
Recently,
the third and the forth pyrochlore superconductors, namely
RbOs$_2$O$_6$ ($T_c\simeq6.3$~K)
\cite{Hiroi04,Bruhwiller04,Kazakov04,Yonezawa04a} and
CsOs$_2$O$_6$ ($T_c\simeq3.3$~K) \cite{Yonezawa04b} were
announced.
The nature of the pairing mechanism in these pyrochlore compounds
is still an open question.
CdRe$_2$O$_7$ is suggested to be a weak--coupling BCS
superconductor \cite{Hiroi02} without
nodes in the superconducting gap \cite{Hiroi02,Lumsden02}.
Specific heat measurements performed on RbOs$_2$O$_6$
\cite{Bruhwiller04} are consistent with BCS type of behaviour.
On the other hand,
Hiroi {\em et al.} \cite{Hiroi04} and Koda {\em et al.}
\cite{Koda04}
pointed to an unconventional mechanism of
superconductivity
in  KOs$_2$O$_6$
and suggested
that the superconducting order parameter is anisotropic
\cite{Koda04}.

Magnetic field penetration depth $\lambda$ and high-pressure
studies traditionally play an important role in superconductivity.
The temperature dependence of $\lambda$ reflects the quasiparticle
density of states available for thermal excitations and therefore
probes the superconducting gap structure. In addition, the
shape
of $\lambda(T)$
can provide
relevant information about the superconducting mechanism.
High--pressure experiments, if a high pressure effect on $T_c$ is
observed, are a good indication that higher values of $T_c$ in
similar compounds may be obtained by "chemical" pressure (by
changing the appropriate ion to its chemical equivalent with
different ion size).

In this letter we report studies of the hydrostatic pressure
effect on the superconducting temperature $T_c$ and the magnetic
field penetration depth $\lambda$ in the pyrochlore
superconductor RbOs$_2$O$_6$. The value of $\lambda$ extrapolated to zero
temperature
is estimated to be in the range 410~nm to 520~nm. The
behavior
of $\lambda(T)$
indicates that RbOs$_2$O$_6$ is most probably a
s--wave
weak--coupled BCS superconductor
within
the adiabatic limit.
However,
d--wave type
of pairing symmetry
is not completely excluded. The transition
temperature increases
with increasing pressure with the slope ${\rm d}T_c /{\rm
d}p=0.090(1)$~K/kbar. This effect can be explained by a
substantial increase of the electron-phonon coupling constant
$\lambda_{el-ph}$ with pressure.

\begin{figure}[htb]
\includegraphics[width=0.9\linewidth]{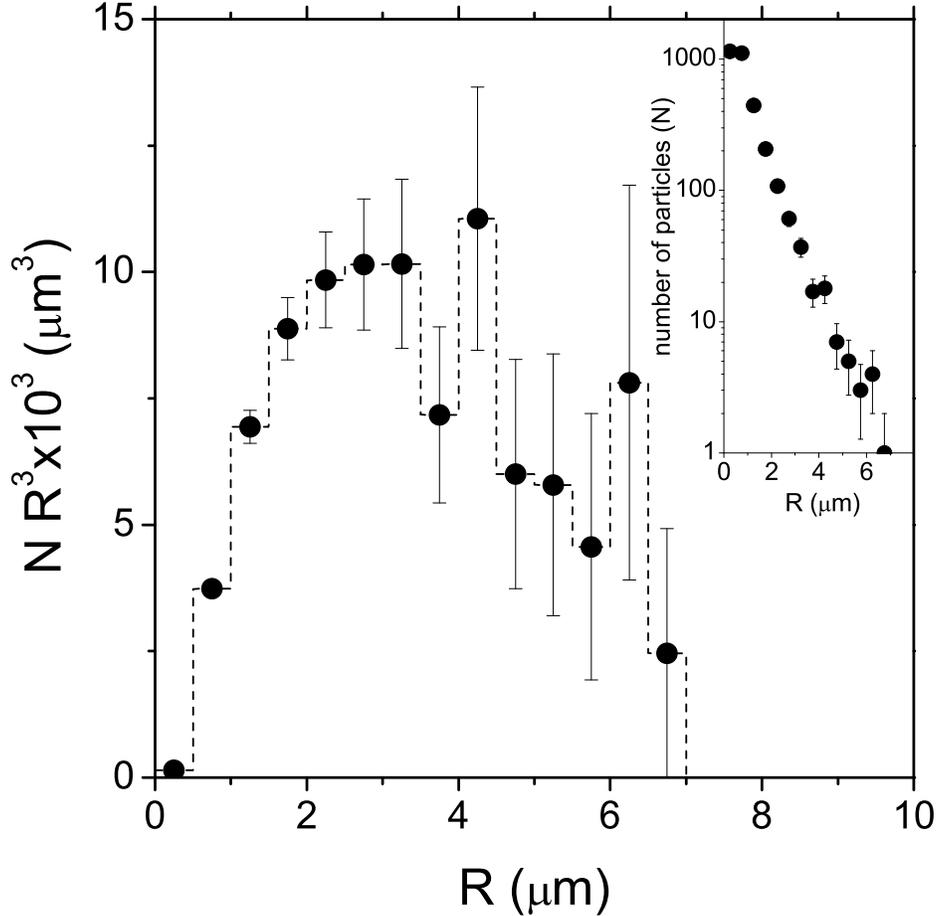}
\caption{ The volume fraction distribution $N(R) R^3$ in the
RbOs$_2$O$_6$ powder determined from the SEM photographs.
The dashed line is the stepwise $g(R)$ function used for
$\lambda(T)$ determination by means of
Eq.~(\ref{eq:Shoenberg-modified}). Inset shows the grain size
distribution $N(R)$ in a semilogarithmic scale. Errors are
statistical.}
 \label{fig:Grain-distribution}
\end{figure}

Details of the sample preparation for RbOs$_2$O$_6$ can be found
elsewhere \cite{Kazakov04,Bruhwiller04}. In the current work, we
performed DC--magnetization measurements. In this technique, the
critical temperature is directly obtained from the magnetization
curve. Following the classical work of Shoenberg
\cite{Shoenberg40}, for
fine powders with known grain sizes the magnetic penetration depth
can be calculated from the Meissner fraction. For this reason
the RbOs$_2$O$_6$ powder sample was ground. The grain size
distribution
was then determined by analyzing SEM
(scanning electron microscope) photographs. The measured particle
radius distribution $N(R)$ and the distribution of the volume
fraction $\sim N(R) R^3$ are shown in
Fig.~\ref{fig:Grain-distribution}.

The hydrostatic pressure was generated in a copper-beryllium
piston cylinder clamp
especially designed for magnetization measurements under pressure
\cite{Straessle02}. The sample was mixed with Fluorient FC77
(pressure transmitting medium) with a sample to liquid volume
ratio
of
approximately $1/6$. With this cell hydrostatic pressures up to
12~kbar can be achieved \cite{Straessle02}. The pressure was taken
from
a separate calibration
set of magnetization experiments where a small piece of In
[$T_{c}(0)=3.4$~K] with known $T_{c}(p)$ dependence was added to
the sample and both $T_c$ of In and RbOs$_2$O$_6$ were recorded.

The field-cooled (FC) magnetization measurements were performed
with a
SQUID magnetometer in a field of $ 0.5$~mT at temperatures ranging
from $1.75$~K to $10$~K. The absence of weak links between grains
was confirmed by the linear magnetic field dependence of the FC
magnetization, measured at $0.25$~mT, $0.5$~mT and $1.0$~mT for
each pressure at $T=1.75$~K. The Meissner fraction $f$ was
calculated from the mass of the samples, their x-ray density, and
assuming spherical grains. The volume of the superconducting
fraction was taken 77 \% in accordance with the heat capacity
measurements performed with this sample \cite{Bruhwiller04}.

\begin{figure}[htb]
\includegraphics[width=0.9\linewidth]{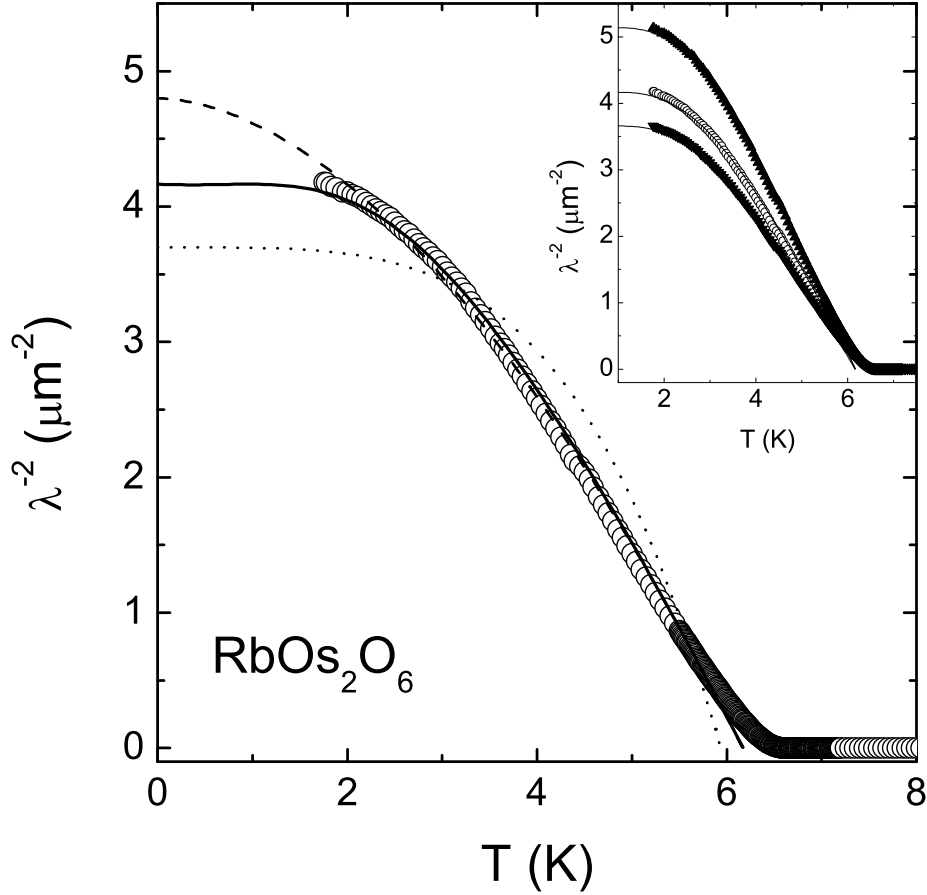}
\caption{The temperature dependence of $\lambda_{ab}^{-2}$
for RbOs$_2$O$_6$
calculated from the measured $f(T)$ by using
Eq.~(\ref{eq:Shoenberg-modified}). Lines represent fit with the
weak (solid line), strong (dotted line) coupling BCS models, and
with a power law (dashed line). See text for an explanation. Inset shows
$\lambda_{ab}^{-2}(T)$ dependences calculated for different
grain--size distribution functions. From top to the bottom:
$N^+(R)$, $N(R)$ and $N^{-}(R)$. }
 \label{fig:Lambda-error}
\end{figure}

The temperature dependence of the penetration depth $\lambda$
was calculated from the measured $f(T)$ by using
the Shoenberg formula \cite{Shoenberg40} modified for the known
grain size distribution \cite{Porch93}:
\begin{eqnarray}
f(T) & = & \int_0^\infty\left(1-
\frac{3\lambda(T)}{R}\coth\frac{R}{\lambda(T)}+
\frac{3\lambda^2(T)}{R^2}\right)g(R){\rm d}R / \nonumber \\
& &\int_0^\infty g(R){\rm d}R \ \ \ .
 \label{eq:Shoenberg-modified}
\end{eqnarray}
Here $g(R)= N(R) R^3$, and $N(R)$ is the grain size distribution
(see Fig.~\ref{fig:Grain-distribution}).
By solving this nonlinear equation, $\lambda$ for each value of
$f$ was extracted, and then the set of $\lambda(T,p)$ dependences
was reconstructed. The resulting temperature dependence
$\lambda^{-2}(T,0)$ at ambient pressure is shown in
Fig.~\ref{fig:Lambda-error}. Reconstructed data were fitted with
different models.
The dotted line represents the fit with
the two--fluid
model $\lambda^{-2}(T)/\lambda^{-2}(0)=1- (T/T_{c})^4$
($T_c=5.91(2)$~K, $\lambda(0)=520(5)$~nm) which corresponds to
a strong coupled BCS
superconductor.
In this paragraph, the errors in parameters are transfered
from the "noise" of magnetization measurements and do not include
systematic errors which are discussed later.
The solid line is the fit ($T_c=6.16(1)$~K,
$\lambda(0)=490(1)$~nm) with the tabulated M\"ulhschlegel data
\cite{Muhlschlegel59} calculated for a weak coupled s-wave BCS
superconductor. It is seen that the weak coupling BCS behavior
describes the experimental data rather well, below 5.9~K the
deviation of the experimental points from the
theoretical BCS curve does not
exceed 2\%. For comparison with
literature,
we also
performed
a
fit with the empirical power law
$\lambda^{-2}(T)/\lambda^{-2}(0)=1- (T/T_{c})^n$
\cite{Zimmermann95} with
free $n$ (dashed line in Fig.~\ref{fig:Lambda-error}). The fit
yields $T_c=6.17(1)$, $\lambda(0)=456(3)$~nm, and $n=1.80(3)$.
In \cite{Koda04}  an
observation of $n\simeq2$ was
taken as an argument in favor of
d--wave type of pairing.
To distinguish between weak coupling BCS and d--wave models
one has
to
know $\lambda^{-2}(T)$
{\it at low temperatures} where
they
exhibit completely different behavior. Unfortunately,
these data
are
not available yet. That is why, from the current experimental data, we
can not exclude completely the possibility of $d_{x^2-y^2}$ type
of pairing in  RbOs$_2$O$_6$. We plan to perform
low temperature measurements
in the nearest future.

To estimate the error in the {\it absolute} value of $\lambda(0)$
introduced by the uncertainty in the grain--size distribution we
performed reconstructions with two "extreme" conditions. For the
first we took the grain--size distribution in the form $N^{-}(R) =
N(R)-\sqrt{N(R)}$, for the second we took $N^{+}(R) =
N(R)+\sqrt{N(R)}$. Appropriate $\lambda^{-2}(T)$ dependences for
$N^+(R)$, $N(R)$ and $N^{-}(R)$ are shown in the insert of
Fig.~\ref{fig:Lambda-error}. The fit with the weak--coupling BCS
model yields 440(1)~nm for the lowest and 520(1)~nm for the highest
values of $\lambda(0)$. Bearing in mind the absence of the
experimental data at low temperatures and, as a result, the model
dependent (power law vs weak--coupling) error in $\lambda(0)$
extrapolation procedure (which introduces additional uncertainty
of about 30~nm)
we can estimate the interval for $\lambda(0)$ ranging from 410~nm
to 520~nm. To diminish this uncertainty more direct measurements
(e.g. $\mu$SR) are planed.

As a next step we performed pressure effect (PE) measurements
on $T_c$ and $\lambda$. As it was already mentioned, the procedure
of the $\lambda^{-2}(T)$ reconstruction is sensitive to the
grain--size distribution. In addition it is also very sensitive to
the value of the superconducting fraction, which was fixed from
heat capacity measurements. The good thing here is that we study
{\it relative} effects measured with the same sample in the same
pressure cell, where most of
the
systematic errors are
eliminated. The
main systematic error for such measurements comes
from misalignments of the experimental setup after the cell was
removed
from the SQUID magnetometer and put
back again.
We checked this procedure with a set of measurements at constant
pressure. The systematic scattering of the magnetization data
is of about $0.5\%$,
giving
a relative error in $\lambda^{-2}(T)$ of about 2\%.

\begin{figure}[htb]
\includegraphics[width=0.9\linewidth]{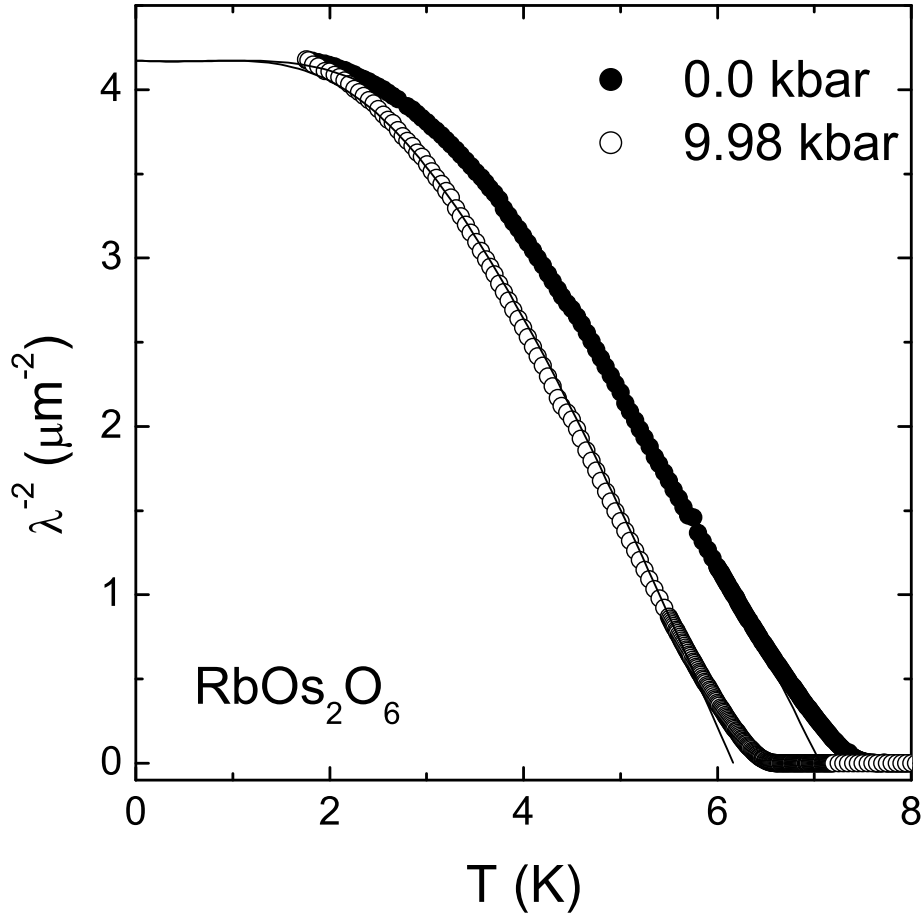}
\caption{ The temperature dependence of $\lambda_{ab}^{-2}$ for
RbOs$_2$O$_6$
obtained from $f(T)$ data at
$p=0.0$~kbar and 9.98~kbar
using Eq.~(\ref{eq:Shoenberg-modified}). Solid lines represent fits with
the weak--coupling BCS model. }
 \label{fig:Lambda-pressure}
\end{figure}

\begin{figure}[htb]
\includegraphics[width=0.9\linewidth]{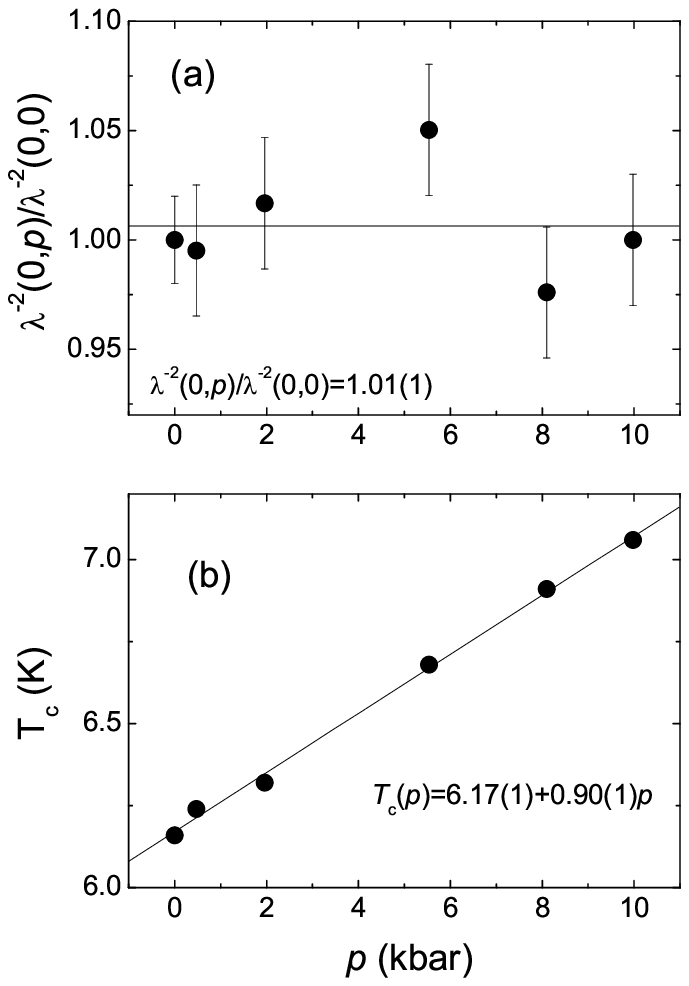}
\caption{
The pressure dependence of
$\lambda^{-2}(0)(p)/\lambda^{-2}(0)(p=0)$ (a) and $T_c$ (b) in
RbOs$_2$O$_6$. The solid lines
are fits with
parameters shown in the figure.}
 \label{fig:Correlations}
\end{figure}

Figure~\ref{fig:Lambda-pressure}  shows the $\lambda^{-2}$ vs $T$
dependences for $p=$0.0~kbar and 9.98~kbar. The transition
temperature increases almost by 1~K at 9.98~kbar, whereas there is
practically no change in $\lambda(0)$. The reconstructed curves
are indistinguishable within the error bars after $T/T_c$ scaling.
This feature implies that $\lambda(0)$ is pressure independent
with the absolute value of $\lambda(0)$ depending on the pairing
model. To reduce the model dependent uncertainties, the normalized
to ambient pressure magnetic penetration depth
$\lambda^{-2}(0,p)/\lambda^{-2}(0,0)$
are
plotted in Fig.~\ref{fig:Correlations}.
One can see that $\lambda^{-2}(0,p)/\lambda^{-2}(0,0)$ data are
scattered and touching by error bars [see
Fig.~\ref{fig:Correlations}(a)].

In RbOs$_2$O$_6$ an estimate of the coherence length $\xi$,
derived from the second critical field $H_{c2}(0)$, gives $\xi
\simeq 7.4 $~nm \cite{Bruhwiller04}.
Under the assumption
that $l$ has the same order of magnitude as in the pyrochlore
superconductor Cd$_2$Re$_2$O$_7$ ($l\sim 20-70$~nm \cite{Hiroi02}),
RbOs$_2$O$_6$
may
be considered as a superconductor in the clean
limit $l\gg\xi$. In this case $\lambda$ obeys the relation:
\begin{equation}
\lambda\propto n_s/m^\ast \ ,
 \label{eq:lambda}
\end{equation}
where  $n_{s}$ is the superconducting charge carrier density, and
$m^{\ast}$ is the effective mass of
the supercarriers. Therefore, the absence of a PE on $\lambda(0)$
suggests that either both quantities $n_s$ and $m^\ast$ are
pressure independent or the pressure shifts of $n_s$ 
and $m^\ast$ cancel each other.
While we cannot
rule out completely the second scenario, we think that the the
first one is more
likely.
The conventional phonon-mediated theory of superconductivity is
based on the Migdal adiabatic approximation in which the effective
supercarrier mass $m^{\ast}$ is independent of the lattice degrees
of freedom. Thus, the absence of a PE on $\lambda(0)$ [see
Fig~\ref{fig:Correlations}(a)] suggests that RbOs$_2$O$_6$
may be considered as an
adiabatic BCS superconductor.
Note that the
same effect (absence of a PE on $\lambda$) was observed recently in
MgB$_2$ which is
accepted to be a purely phonon mediated superconductor
\cite{Dicastro_unp}.

The pressure dependence of the critical temperature is shown in
Fig.~\ref{fig:Correlations}(b). 
The linear fit yields ${\rm d}T_c/{\rm d}p=0.090(1)$~K/kbar. The
linear increase of $T_c$ with pressure observed in RbOs$_2$O$_6$
is quite unusual. For the majority of BCS--type superconductors
(including MgB$_2$) $T_c$ decreases with increasing pressure.
For
conventional
superconductors the pressure shift
of $T_c$ can be derived as \cite{McMillan68,Hodder69}
\begin{equation}
\frac{{\rm d}\ln T_c}{{\rm d}p}= \frac{{\rm d}\ln
\langle\omega\rangle}{{\rm d}p}+
\frac{1.23\lambda_{el-ph}}{(\lambda_{el-ph}-0.11)^2}\frac{{\rm d}
\ln \lambda_{el-ph}}{{\rm d}p}\ ,
 \label{eq:delta_Tc}
\end{equation}
where $\langle\omega\rangle$ is the average phonon frequency,
and $\lambda_{el-ph}$ is the
electron--phonon coupling constant.
There are two contributions to the pressure shift of $T_c$: from
the phonon system  [${\rm d}\ln \langle\omega\rangle/{\rm d}p$]
and from the coupling between electron and phonon subsystems
[${\rm d} \ln \lambda_{el-ph}/{\rm d}p$]. The effect of pressure
on
the
phonon spectra usually results in an increase of the
average phonon frequency,
and the first term in
Eq.~(\ref{eq:delta_Tc}) is generally positive.
An estimate of a typical range of ${\rm d}\langle\omega\rangle /
{\rm d}p = -\gamma/B$ (where $\gamma={\rm
d}\ln\langle\omega\rangle/{\rm d}\ln V$ is the Gr\"uneisen
parameter, $V$ is the volume, and $B$ is the bulk modulus) in
conventional superconductors gives ${\rm d}\langle\omega\rangle /
{\rm d}p\approx 0.01\%-0.5\%$ per kbar
\cite{Seiden69,WebElements}.
The pressure
shift of $T_c$
found in our study
(${\rm d} \ln T_c/{\rm d}p\simeq1.5\%/$kbar) is much bigger than
the possible contribution form the first term in
Eq.~(\ref{eq:delta_Tc}). Therefore,
our results
suggest a substantial increase of
$\lambda_{el-ph}$ with pressure.
Alternatively, the positive pressure effect on
$T_c$ can be explained by a sort of charge ordering, resulting
in the disproportionation of the Os tetrahedra \cite{Hiroi02b}.

In conclusion, we performed magnetization measurements in the
newly discovered superconductor RbOs$_2$O$_6$ under hydrostatic
pressure. The absolute value of $\lambda$ at zero temperature and
ambient pressure is estimated to be in the range 410--520~nm. The
temperature dependence of the magnetic penetration depth $\lambda$
is consistent with that expected for a weak--coupled s--wave BCS
superconductor. However, to rule out completely d--wave symmetry,
additional low temperatures measurements are required. A
pronounced and {\it positive} pressure effect  on $T_c$ with ${\rm
d} T_c/{\rm d} p =0.090(1)$~K/kbar was observed, in contrast to
the negative pressure shift generally detected in conventional
superconductors. This finding can be explained within the
framework of BCS theory under the assumption that the
electron-phonon coupling constant $\lambda_{el-ph}$
increases with pressure. The absence (within the experimental
uncertainties) of the pressure effect on $\lambda$ suggests that
RbOs$_2$O$_6$ is an adiabatic BCS--type superconductor.

This work was supported by the Swiss National Science Foundation
and by the NCCR program \textit{Materials with Novel Electronic
Properties} (MaNEP) sponsored by the Swiss National Science
Foundation.

\end{document}